\documentclass[lettersize,journal]{IEEEtran} 

\usepackage{xcolor}
\usepackage[inkscapelatex=false]{svg}
\usepackage{url}

\usepackage{graphicx}
\usepackage{svg}
\usepackage{adjustbox}
\usepackage{algorithm}
\usepackage{algpseudocode}
\usepackage[font={footnotesize}]{caption}
\usepackage[font={footnotesize}]{subcaption}
\usepackage{graphicx}
\usepackage{comment}
\usepackage{float}
\usepackage{mathtools}

\algnewcommand{\algorithmicand}{\textbf{ and }}
\algnewcommand{\algorithmicor}{\textbf{ or }}
\algnewcommand{\OR}{\algorithmicor}
\algnewcommand{\AND}{\algorithmicand}

\begin{document}

\title{ Autonomous Hyperspectral Characterisation Station: \\Robotically Assisted Characterisation of Polymer Degradation}

\author{Shayan Azizi$^{1,2}$, Ehsan Asadi$^{1}$, Shaun Howard$^{2}$, Benjamin W. Muir$^{2}$, Riley O'Shea$^{2}$ and Alireza Bab-Hadiashar$^{1}$%
\thanks{$^{1}$Shayan Azizi, Ehsan Asadi and Alireza Bab-Hadiasher are with Faculty of Manufacturing, Materials and Mechatronics Engineering, RMIT University, VIC 3000, Australia 
        (email: s3950426@student.rmit.edu.au; ehsan.asadi@rmit.edu.au; alireza.bab-hadiashar@rmit.edu.au)}%
\thanks{$^{2}$Shayan Azizi, Shaun Howard, Benjamin W. Muir and Riley O'Shea are with the Manufacturing business unit at CSIRO, VIC 3168, Australia
        (email: shayan.azizi@csiro.au; shaun.howard@csiro.au; ben.muir@csiro.au; riley.o'shea@csiro.au)}%
}

%%%%%%%%%%%%%%%%%%%%%%%%%%%%%%%%%%%%%%%%%%%%%%%%%%%%%%%%%%%%%%%%%%%%%%%%%%%%%%%%
\maketitle

\begin{abstract}

This paper addresses the gap between the capabilities and utilisation of robotics and automation in laboratory settings and builds upon the concept of Self Driving Labs (SDL). %to significantly impact laboratory operations. 
We introduce an innovative approach to the temporal characterisation of materials. The article discusses the challenges posed by manual methods involving established laboratory equipment and presents an automated hyperspectral characterisation station. This station integrates robot-aided hyperspectral imaging, complex material characterisation modeling, and automated data analysis, offering a non-destructive and comprehensive approach. This work explains how the proposed assembly can automatically measure the half-life of biodegradable polymers with higher throughput and accuracy than manual methods. The investigation explores the effect of pH, number of average molecular weight (Mn), end groups, and blends on the degradation rate of polylactic acid (PLA). The contributions of the paper lie in introducing an adaptable classification station for novel characterisation methods and presenting an innovative methodology for polymer degradation rate measurement. The proposed system has the potential to accelerate the development of high-throughput screening and characterisation methods in material and chemistry laboratories.

\end{abstract}

\def\abstractname{Note to Practitioners}
\begin{abstract}
The characterisation and classification of materials hold significant importance within the realms of material science, chemistry, manufacturing and circular economy. We introduce an innovative approach by employing robotics to automate hyperspectral imaging and processing to accelerate and improve material characterisation within a laboratory environment. This methodology facilitates the time-dependent characterisation of materials. This automated system encompasses sample manipulation and handling for hyperspectral scanning, accompanied by automated image and data processing procedures to minimise manual interventions effectively. The developed system can be seamlessly integrated into lab settings, offering objective, accurate, and automated measurement of biodegradable polymer degradation rate.
\end{abstract}


%%%%%%%%%%%%%%%%%%%%%%%%%%%%%%%%%%%%%%%%%%%%%%%%%%%%%%%%%%%%%%%%%%%%%%%%%%%%%%%%
\section{INTRODUCTION}

\begin{figure}[t]
    \centering
    \includegraphics[width=1\columnwidth]{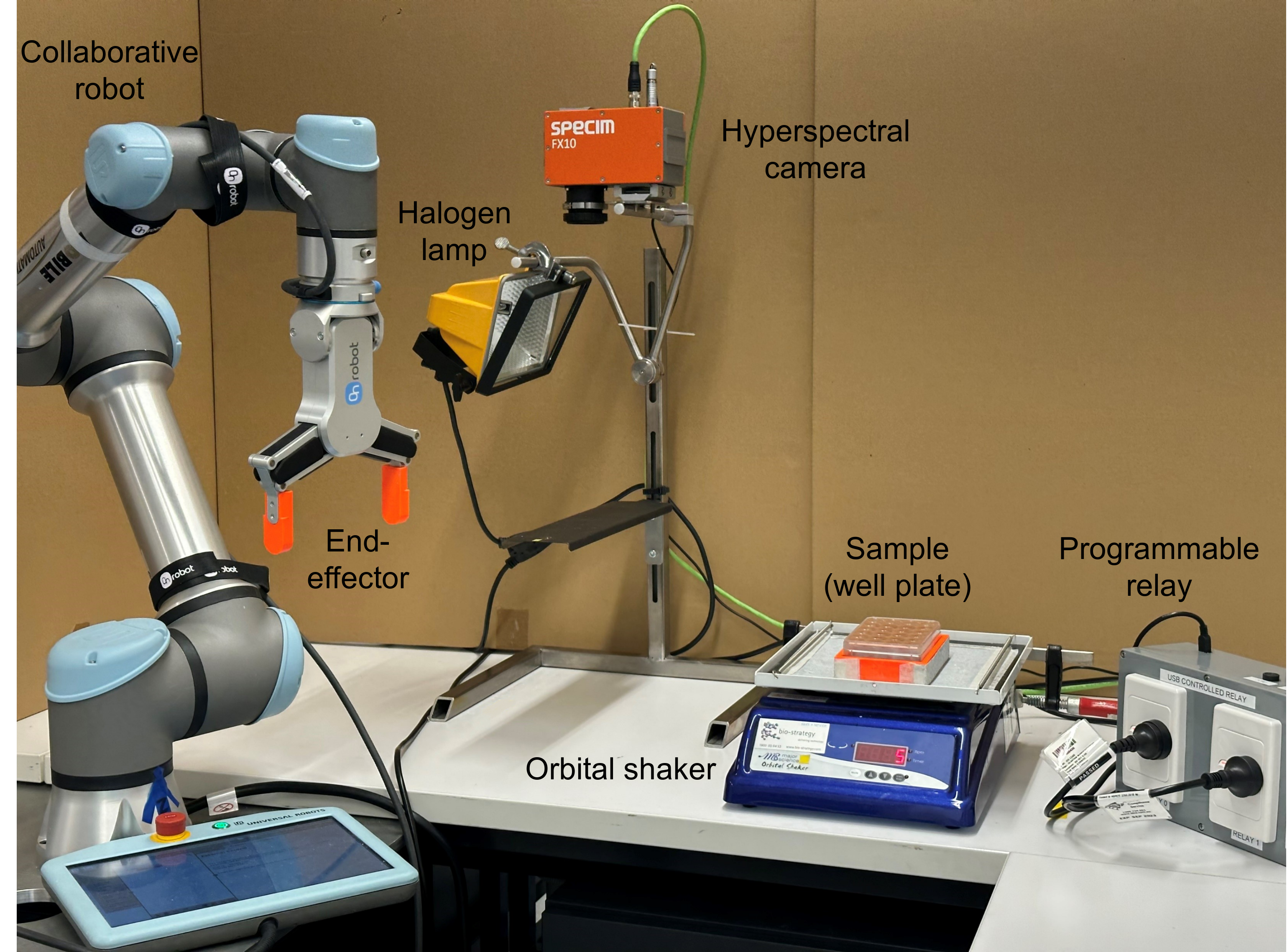}
    \caption{Robot-aided hyperspectral mapping and data sampling. \vspace{-0.5cm}}
    \label{fig: setup_image}
\end{figure}

\IEEEPARstart{E}{fforts} have been directed towards bridging the gap between capabilities and utilisation of robotics and automation within a laboratory setting. With the introduction of Self Driving Labs (SDL) \cite{abolhasani2023rise} such as the mobile chemist \cite{burger2020mobile}, automation and AI are expected to significantly impact laboratory operations, enabling mechanised large-scale experimentation with high levels of repeatability and throughput while reducing operational costs \cite{prabhu2017dawn}. Automation can greatly enhance material classification and characterisation which are essential tasks within scientific laboratories.
Specifically, there has been a growing interest in the use and applications of biodegradable polymers, including but not limited to packaging, electronics, automotive components, and critical medical applications such as implantation and drug delivery \cite{samir2022recent}\cite{haider2019plastics}. Consequently, automation of degradation rate measurement and analysis is becoming increasingly important, necessitating objective data sampling throughout days, nights and weekends for an extended period of time.

\subsection{Problem, solution and rationale}
Numerous chemical and material science experiments involve reactions or processes that happen gradually, yielding crucial data at specific time points. Prolonged experiments that require regular measurements impose significant demands on laboratory staff and equipment usage. Moreover, many experiments prove sensitive to intrusive measurement techniques entailing resampling or extreme exposure. Established equipment for characterisation, including mass spectrometry, Fourier-transform infrared spectroscopy (FTIR), nuclear magnetic resonance (NMR), UV-vis spectroscopy, gas permeation chromatography (GPC), etc., are frequently operated manually, requiring laborious sample preparation. They often have prolonged operation times and are associated with high operational and setup costs. These characteristics limit their seamless integration into automated workflows.

Current techniques for measuring the degradation rate of biodegradable polymers encompass a range of methods, including NMR, FTIR, GPC, thermogravimetric analysis (TGA), (SEM) and turbidity measurements \cite{baidurah2022methods}. For instance, GPC quantifies alterations in the hydrodynamic radius associated with molecular weight. By evaluating changes in molecular weight, insights into the polymer degradation rate can be derived. However, GPC involves many preparatory manual steps as well as high levels of uncertainty associated with its measurements. Alternatively, a turbidity scanner can measure the intensity of scattered light by the sample, which may decrease as the polymer degrades. Turbidity measurements are based on assumptions that the polymer is homogeneous within the solution and the light beam wavelength is appropriate for all samples, which is usually not the case. These standard analytical techniques exhibit limitations in assessing degradation rates and often involve multiple manual procedures with low throughput.

This paper presents an automated hyperspectral characterisation station for high-throughput material analysis without requiring any preparatory steps or deep domain knowledge. This method allows flexibility and optimised automation via integrating robot-aided hyperspectral imaging with complex material characterisation modelling and automated data analysis. The compact and non-destructive nature of the proposed method allows synergistic deployment with various classification tools, thereby providing a comprehensive analysis. Through a detailed case study, we show how the proposed assembly can automatically measure the half-life of bio-degradable polymers with higher throughput and accuracy compared to what is typically achieved with manual methods.

\subsection{Contributions and capabilities}
This paper has two major contributions: \textbf{(I)} We introduce an adaptable and automated classification station that enables the development of novel characterisation methods involving slow temporal changes by providing readily analysable hyperspectral data over an extended period of time. Applications of this system encompass tracking changes within samples over time, conducting comparative analyses across multiple samples, time-dependent characterisation, and non-intrusive monitoring. The hyperspectral data provides a range of information, including variations in total reflectance, alterations in peak and trough intensities, quantitative colorimetric assessment, and identification of anomalies in established spectral patterns. This information is useful to detect the presence or absence of specific materials as well as to measure solubility and degradation, light sensitivity, uniformity, concentrations, etc. This framework provides opportunities for the development of innovative, rapid, automated, and high-throughput screening and characterisation methods within material and chemistry laboratories. The use of the proposed system has the potential to accelerate the generation of training data for machine learning methods that can solve characterisation challenges. \textbf{(II)} An innovative methodology is developed for polymer degradation rate measurement as an objective, accurate and automated solution with minimum human intervention. This technique provides insights into the degradation of Polylactic Acid (PLA). The impact of variables such as pH, molecular weight, end groups and blends of polymers on the rate of polymer degradation is investigated.

\section{Related work}
Laboratory automation is increasingly necessary to speed up chemical analysis processes and reduce the required human resources \cite{abolhasani2023rise}\cite{prabhu2017dawn}. Recent work are focused on using collaborative robots (cobots), which are an emerging technology that brings flexibility to labs without compromising scientists' safety. As an example, ``the mobile robotic chemist" \cite{burger2020mobile} is designed to perform complete experiments and improve future experiments by analysing the results of previous runs. Such closed-loop automated processes are intended to increase the speed of laboratory processes. A less sophisticated application of cobots in labs is their use in Covid-19 testing facilities \cite{zanchettin2022collaborative}, where a sample cleaning procedure is automated, saving lab technicians time and effort from performing tedious and repetitive tasks. Another example is a robotic system performing the Michael reaction experiment competently as a junior chemist \cite{lim2020development}. 
To add flexibility, a system architecture called the ARChemist \cite{fakhruldeen2022archemist} was developed to facilitate laboratory automation using existing equipment. Similarly, a scene-aware robotic chemist with the ability to plan long horizon trajectories and perform pouring exercises was developed to perform basic solubility and crystallisation experiments \cite{yoshikawa2023chemistry}.

Robots are also increasingly employed to automate various aspects of hyperspectral analysis across different domains. As an example, robots and HSI were used to design a construction demolition waste system \cite{xiao2020development} to separate common waste materials such as brick, rubber, concrete, and wood using a robotic arm. For outdoor applications, an automated hyperspectral imaging system, in conjunction with a deep learning technique, was also developed to predict the nitrogen contents in corn plant leaves \cite{chen2021automated}. However, automated systems have turned out to have a lower time-averaged throughput compared to humans. Most recently, a system known as Hyperbot-A \cite{hanson2022hyperbot} has been developed to automate the process of hyperspectral imaging data gathering with the help of a cobot. However, hyperbot-A’s integration with external equipment remains unexplored and hasn't been utilised for specific applications. 

In terms of data analysis, there is significant potential to improve scientific measurements by using artificial intelligence, particularly those focused on hyperspectral imaging (HSI)    \cite{sagan2021data}. For instance, a study conducted in a laboratory setting utilises HSI to predict the mean size of polymer molecules \cite{pieszczek2021near}, which had an observed and predicted value correlation (R2) of 0.97. Similarly, by using machine learning and hyperspectral images, the possibility of measuring the conductivity of thin polymer films was explored \cite{bash2020machine}. Although conductivity and reflection spectrum may initially seem unrelated, the combination of HSI and ML served as a diagnostic tool to measure film conductivity. While this work demonstrated the potential of hyperspectral imaging as a characterisation technique, the data-gathering process was conducted manually.

\begin{figure*}[t]
    \centering
    \includegraphics[width=2\columnwidth]{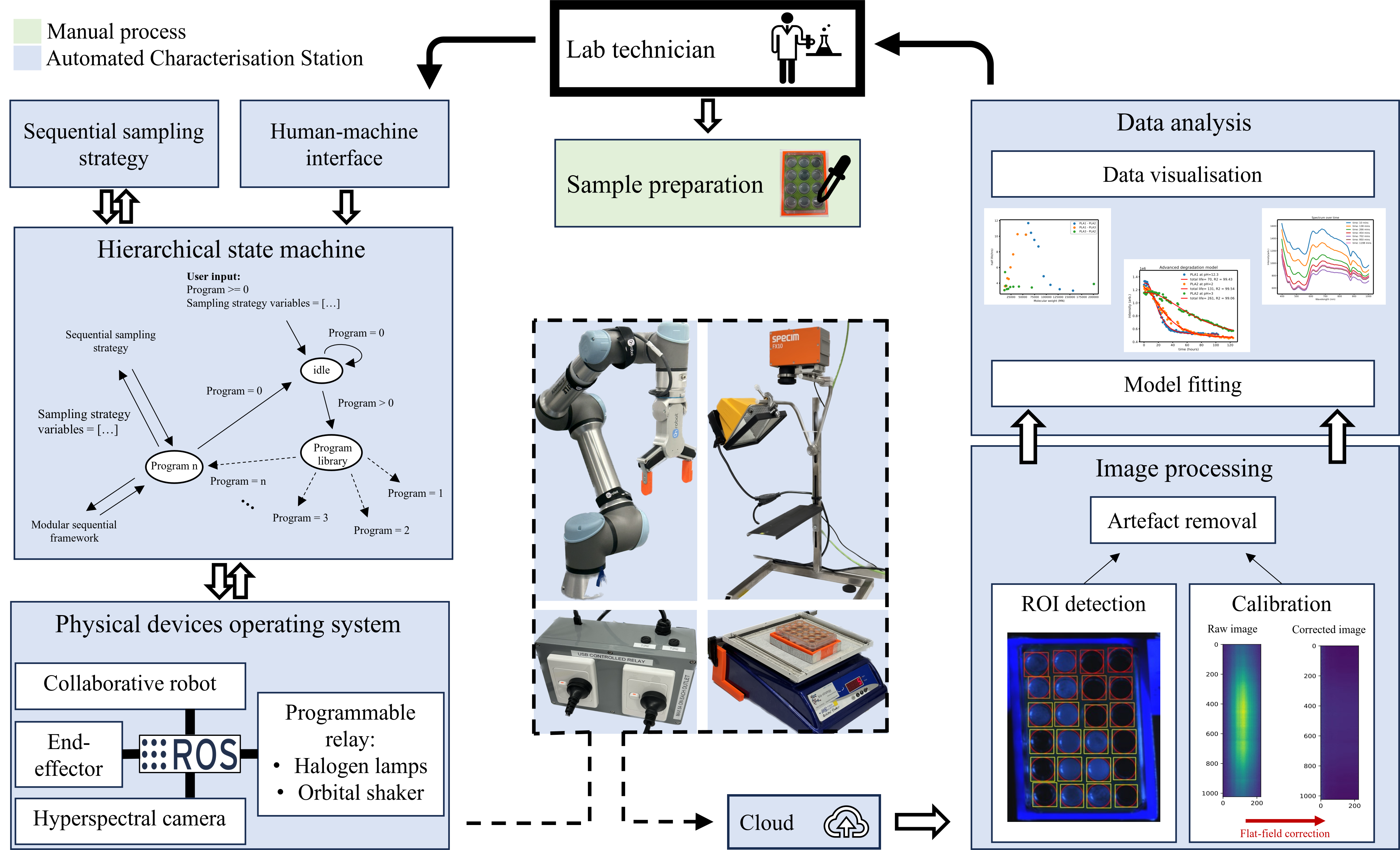}
    \caption{Overall system design of autonomous hyperspectral characterisation station for laboratories.}
    \label{fig: overall system}
\end{figure*}

Over the past few years, some work has also investigated polymer degradation and aging using hyperspectral imaging. Polylactic acid degradation was evaluated in an accelerated weathering test using a NIR hyperspectral camera, which used spectral intensity to suggest a decrease in molecular weight induced by the hydrolysis degradation \cite{shinzawa2012accelerated}. Another study of polymer degradation \cite{dorrepaal2018hyperspectral} found that the HSI method provides a more detailed picture of sample degradation in terms of its spatial variance. Polymer damage and aging were also analysed for silicone rubbers while mainly utilising the visible spectrum \cite{bleszynski2020visualizing}. Most recently, the aging state of styrene–butadiene–styrene (SBS) has been analysed using a Short Wave Infrared HSI, predicting the aging state with an error of 1.8-2.1 days \cite{wieser2022application}. None of the mentioned work examines polymer degradation in a chemical lab that allows fine-tuning of different variables. Our proposed methodology is a generalised approach to characterising and quantifying the degradation rate of various polymers due to hydrolysis. It provides a methodology for analysing a larger set of applications than scenario-specific methods and measurements.

\section{Automated Hyperspectral Characterisation Station}
This section provides an overview of the innovative characterisation station proposed as a key element of process automation within chemistry laboratories, enabling outcomes with objective, data-driven and quantifiable results. Enabled by the capabilities of the proposed system, in section \ref{polydeg_method}, we introduce a new and fully automated approach for measuring and analysing polymer degradation rates.
The automation station is designed around multiple factors such as safety, flexibility, minimising human intervention and maximising throughput. The system consists of two primary processes: the characterisation station and the image and data analysis pipeline. The station is comprised of five principal components: a manipulator (collaborative robot), an end-effector (gripper), an imaging device (hyperspectral camera), illumination sources (halogen lamps), and lab equipment (shaker). These components are integrated through the Robotic Operating System (ROS) as the communication framework. The data and imaging analysis pipeline comprises five key sub-processes: calibration, region of interest (ROI) detection, artefact removal, model fitting, and data visualisation.

\subsection{Process automation and hierarchical state machine}
In our setup, we developed a series of independent software modules that each controlled a physical device. The user can arrange the modules in any sequence to define the operations required for a task. The resulting sequential state machine is called a ``program" and can be saved into a library for future use. Combining several programs and a user input module comprises a hierarchical state machine that enables the user to define a sequence of controlled actions as shown in Fig. \ref{fig: overall system}. ``Program" number refers to the labelling associated with individual subtasks, and ``sequential sampling variables'' serves as the starting point for the sampling strategy, determining the number of iterations a program runs and specifying the time interval between each iteration.. Fig. \ref{fig: img_proc} shows an example program consisting of modular components for the automation of an imaging task. Each program performs a specific task via its own state machine, and the human-machine interface allows for serialised commands, orchestrating multiple programs. Such tasks include running experiments in parallel or performing more complex characterisation of high-level tasks.

\subsection{Image processing and data analysis}
Hyperspectral images are dense and high-dimensional data structures that require the implementation of various image and data processing techniques to enable meaningful analysis. Several techniques are employed in our system, including flat field correction, artefact removal and ROI detection. A brief overview of these techniques is as follows.

\textbf{Calibration}: The well-known flat-field correction was employed to account for the influence of lighting on the samples. A highly reflective (99\%) tile was utilised to measure the light source's spectrum (W). The background noise (dark) spectrum (D) was recorded with the shutter closed. Subsequently, the hyperspectral data is calibrated using $C = \frac{R - D}{W - D} \cdot m$, where $R$ and $C$ are the raw calibrated data, respectively and $m$ is a constant scaling factor.

\textbf{ROI detection}: Numerous containers commonly found in laboratory settings, including well plates, petri dishes, and vials, exhibit circular geometries. OpenCV Hough’s circle detection algorithm \cite{opencv_library} with additional developed optimisations and fine-tuning were utilised to automate the identification of regions of interest (ROIs) \cite{Dong2018FastED}\cite{Dong2019RealTimeRM}. The algorithm guarantees non-overlapping detection of circles and sequential numbering, ensuring the convenient identification of each individual sample. Particular focus was placed on the detection of well plates as it allows for concurrent imaging of multiple samples.

\textbf{Artefact removal}: Two primary artefacts are encountered in the detected regions of interest. Firstly, there is glare on the surface of the solution, which is considerably more reflective than the solution, and secondly, there is a possibility of regions being erroneously flagged as part of the sample post-circle detection. To address both concerns, a reflective mask is produced via 3D printing and fitted on top of the well plate and around the samples. This mask reflects considerably more light, sufficient to induce CMOS sensor saturation. Algorithmically, pixels that experience sensor saturation at any wavelength are excluded from the region of interest, thereby resolving both artefacts and enhancing the overall outcomes.

\textbf{Model fitting}: A major challenge in transitioning from lab-scale chemistry to process automation is the lack of quantitative chemical synthesis and analysis. We utilise a model-fitting algorithm to match a particular model, whether it's a basic exponential function or a more complex one, with the data obtained from hyperspectral imaging for various analyses. For example, our novel approach to resolve this issue and calculate the polymer's half-life using only hyperspectral imaging is outlined in the subsequent section.

\section{Automating Polymer Degradation Rate Measurement and Analysis}
\label{polydeg_method}

\subsection{{Biodegradable polymers and their characteristics}}
In this paper, polymer degradation refers to the breakdown of large polymers into smaller units due to hydrolysis, as seen in Fig. \ref{fig: poly_deg_stages}. The underlying mechanisms governing polymer degradation are complex and not fully understood \cite{LAYCOCK2017144}. Existing research suggests two established approaches: bulk erosion and surface erosion \cite{von2002degradable}\cite{GOPFERICH1996103}. Numerous theoretical models have been proposed to describe polymer degradation rates, adopting various approaches, including diffusion and surface area dependence. This study uses a simple exponential decay approach for assessing polymer degradation. The first-order rate equation for the mass of undissolved polymers ($m_p$) is given by \ref{eq:diff_eq}, where $k_1$ is the decay constant.

\begin{equation}
    \label{eq:diff_eq}
    -\frac{dm_p}{dt} = k_1  m_p.
\end{equation}
Solving the differential equation \ref{eq:diff_eq} results in :
\begin{equation}
    \label{eq:diff_eq_sol}
    m_p(t) = a e^{-k_1 t}.
\end{equation}

This study mainly focused on measuring polylactic acid (PLA) degradation. It’s been shown that PLA degrades to lactic acid and its oligomers \cite{inkinen2011lactic}. Additionally, Lactic acid and lactic acid salts are highly soluble in water \cite{pubchem_lactic_url}\cite{JR9540000550}. Hence, degraded PLA will fully dissolve in potassium buffer saline (PBS) and not reflect any visible light. Under these conditions, potassium lactate would be the major product formed at equilibrium. This is further confirmed by an experiment measuring the light reflected from PBS + lactide, PBS + lactic acid and the PBS on its own. As shown in Fig. \ref{fig: three_controls}, the lactide and lactic acid do not contribute to a signal change. The slight difference in the baseline value of the three samples is due to the position of the samples with respect to the camera and light. 

The proposed methodology applies to a variety of polymers as long as they satisfy two conditions: (\textbf{I}) the polymer is relatively well dispersed within the solution (this technique isn't as accurate for visibly concentrated or agglomerated polymers) and ({\textbf{II}) the polymer dissolves into PBS (or other solution) after degradation. An additional polymer independent condition (\textbf{III}) also needs to be satisfied to ensure the mass of polymer per imaged-surface-area ($\rho A$) doesn't exceed a certain threshold ($\tau$).

\begin{figure}[t]
    \centering
    \includegraphics[width=1\columnwidth]{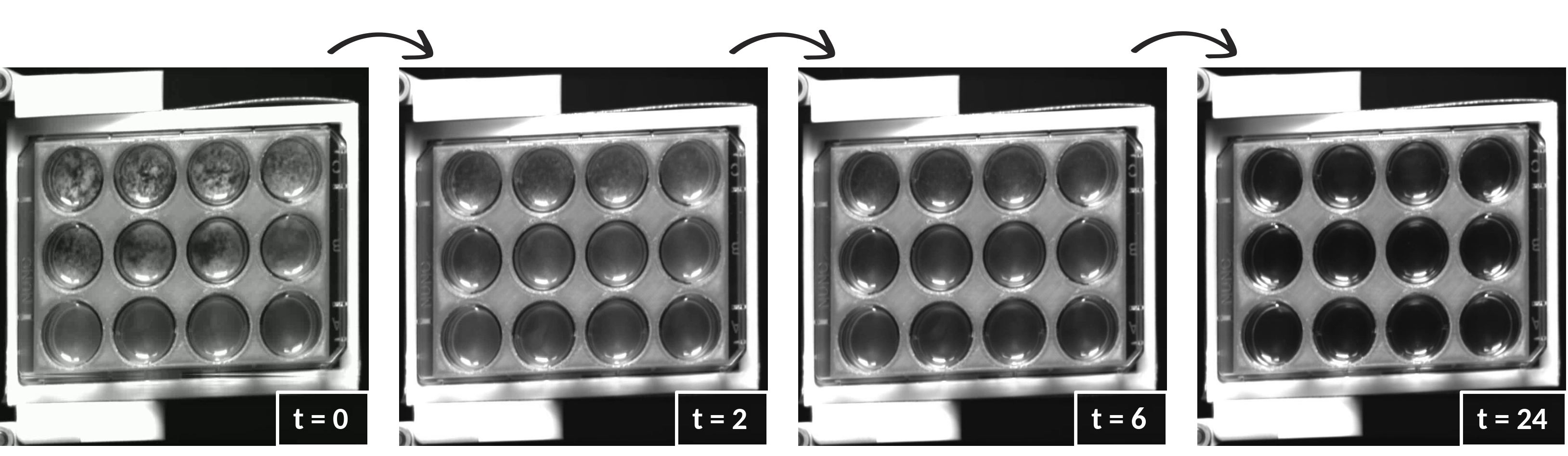}
    \caption{A black and white representation of hyperspectral images, used to measure the degradation rate of polymers. The degradation process is evident as time (in hours) progresses from the left to the right image.}
    \label{fig: poly_deg_stages}
\end{figure}

\subsection{Hyperspectral imaging model for polymers degradation}
Hyperspectral imaging provides measurements of the reflected light over a wide range of wavelengths from biodegradable polymers, in contrast to conventional RGB cameras or similar sensors that are restricted to certain wavelengths. The half-life of polymers can be estimated by fitting an exponential decay model to regular measurements of total reflected light over an extended period of time using a hyperspectral camera.
The total light reflected by a sample is modelled as follows:

\begin{equation}
    \label{eq:reflection_sum}
    R_T = R_s + R_p + R_b,
\end{equation}

where $R_T$ is the total light reflected, $R_s$ is the light reflected by PBS, $R_p$ is the light reflected by any undissolved polymer, and $R_b$ is the light reflected by the background. Since the $R_s$ and $R_b$ remain unchanged over time, as the polymer degrades, they can be replaced by a constant $c$. Hence, we can rewrite Eq. \ref{eq:reflection_sum} as:

\begin{equation}
    \label{eq:simple_reflection}
    R_T = R_p + c
\end{equation}

As previously discussed, we assume that the polymer is relatively well dispersed within the solution and its area density doesn't exceed the threshold $\tau$. This means that the light reflected by the polymer would have a linear relationship with the cumulative amount of undissolved polymer. Notably, the mass of undissolved polymers ($m_p$) maintains direct proportionality with the intensity of reflected light, owing to the fact that non-degraded polymers can't dissolve within the solution. This is important as only the undissolved polymers will contribute to the hyperspectral reflectance signal. As such, we can reasonably expect a linear relationship between reflection and undissolved mass given by:

\begin{equation}
    \label{eq:reflection_mass_relation}
    R_T = a \cdot m_p + c
\end{equation}

\begin{figure}
    \centerline{\includegraphics[width=0.9\columnwidth]{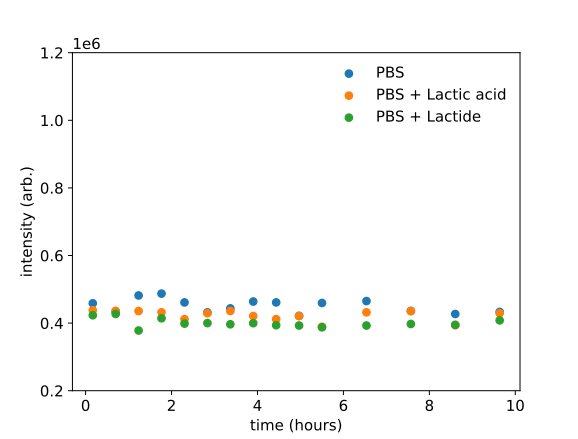}}
    \caption{Total light reflected by the control samples. Lactide and Lactic acid aren't contributing to a signal change (plotted with the same range as the actual results shown in subsequent figures). The standard deviation of reflection intensity values for these controlled samples is less than 5\% of their average values for all cases.}
    \label{fig: three_controls}
\end{figure}

where $a$ and $c$ are both constants. As given in Eq. \ref{eq:diff_eq_sol}, the mass of the remaining polymers can be modelled using an exponential decay, resulting in the following equation: 

\begin{equation}
    \label{eq:final_decay}
    R_T = a \cdot e^{-k_d t} + c
\end{equation}

from which the half-life ($t_\frac{1}{2}$) of the polymer can be calculated by:

\begin{equation}
    \label{eq: dec_to_halflife}
    t_\frac{1}{2} = \frac{\log(2)}{k_d}.
\end{equation}

\subsection{Advanced surface-mass dependent degradation model}
By assuming a spherical shape and constant density ($\rho$) for the polymers, one can arrive at a surface area ($SA$) dependent rate of degradation as given in Eq. \ref{eq: sa_deg_rate}, where $k_d$ is the rate constant \cite{chamas2020degradation}.

\begin{equation}
    \label{eq: sa_deg_rate}
    -\frac{dm_p}{dt} = k_2 \cdot \rho \cdot SA = k_2 (4\pi\rho)^\frac{1}{3}(3m_p)^\frac{2}{3}
\end{equation}

With the linear combination of the above equation and the mass-dependent model given in Eq. \ref{eq:diff_eq}, we derive a new degradation rate model (Eq. \ref{eq:new_model_de}). This incorporates both the surface and bulk modes of degradation into one equation.

\begin{equation}
    \label{eq:new_model_de}
    -\frac{dm_p}{dt} = k_{2} (4\pi\rho)^\frac{1}{3}(3m_p)^\frac{2}{3} + k_{1} \cdot m_p
\end{equation}

Denoting the constant term $k_{2}(4\pi\rho)^\frac{1}{3}(3)^\frac{2}{3}$ by $b$, the differential equation \ref{eq:new_model_de}, will be a Bernoulli ODE that can be solved exactly with solution given by Eq. \ref{eq:new_model_de_solved}.

\begin{equation}
    \label{eq:new_model_de_solved}
    m_p = -\frac{e^{-k_{1} t}(b e^{\frac{k_{1} t}{3}} - e^{d_1 k_{1}})^3}{k_{1}^3}
\end{equation}

By substituting the Eq. \ref{eq:new_model_de_solved} in Eq. \ref{eq:reflection_mass_relation}, we can now model the total reflected light by:

\begin{equation}
    \label{eq:new_model_solved}
    R_T = -\frac{ae^{-k_{1} t}(b e^{\frac{k_1 t}{3}} - e^{d_1 k_{1}})^3}{k_{1}^3} + c
\end{equation}

where $a > 0$, $b \ge 0 $, $k_{1} \ge 0$ and $c \ge (\frac{b}{k})^3$. Variable $a$ is just a scaling factor; for simplicity, we assume $a = 1$. Akaike information criterion (AIC) and Minimum Description Length (MDL) given in equations \ref{eq:AIC} and \ref{eq:MDL} respectively \cite{GHEISSARI20081636}, are used to evaluate the performance of the fitted models.

\begin{equation}
    \label{eq:AIC}
    AIC = \sum_{i=1}^{N}r_{i}^{2} + 2P\sigma^2
\end{equation}

\begin{equation}
    \label{eq:MDL}
    MDL = \sum_{i=1}^{N}r_{i}^{2} + \frac{P}{2}\log(N)\sigma^2
\end{equation}

\subsection{Robot position for imaging}
To calculate the appropriate distance ($d$) and velocity ($V_c$) of a sample with respect to a fixed camera, we need to know the camera's field of view ($FOV$), effective slit width ($L_e$), frame rate ($f$) and spatial resolution ($R_p$) as well as the desired image length ($L$). The desired image length in practice should be slightly larger than the sample width (e.g. by $10\%$). Equations \ref{eq: optical_width} and \ref{eq: arm_speed} show how a sample's distance and velocity values are calculated. 

\begin{equation}
    \label{eq: optical_width}
    % L=2 \cdot d \cdot \tan(\frac{FOV^{\circ}}{2})+L_e,
    d = \frac{L-L_e}{2} \cot(\frac{FOV^{\circ}}{2})
\end{equation}

\begin{equation}
    \label{eq: arm_speed}
    V_c= \frac{L}{R_p} \cdot f,
\end{equation}

\subsection{Sequential sampling strategy with time-varying intervals}

\begin{algorithm}[t]
\caption{Sequential Sampling Strategy}\label{alg: time_opt}
\begin{algorithmic}[1]
\renewcommand{\algorithmicrequire}{\textbf{Input:}}
\renewcommand{\algorithmicensure}{\textbf{Output:}}
\Require{All hyperspectral ROI ($D$), their corresponding time ($t$) and variables $f$, $b_l$, $b_u$, $g$}
\Ensure{$t_{n+1}$, optimal time separation for the next sampling}

\While{sampling}
    \If{new measurement}
        \State $I \gets F(D,t)$ 
    \EndIf

    \State $i_e \gets G(c,J)$ \Comment{Where $J \subseteq I$}
    
    \If{($i_e < \epsilon$)}
        \State Stop Sampling

    \ElsIf{$|i_e - \Delta i| < \delta$}
        \State Continue without change
        
    \Else
        \State $d \gets N(i_e, \Delta i, g)$
        \State $s \gets T(d, f)$
        \State $t_{n+1} \gets P(\Delta t, s, b_l, b_u)$
    \EndIf
\EndWhile

\end{algorithmic}
\end{algorithm}

One of the challenges of polymer degradation analysis lies in determining the optimal time intervals for capturing hyperspectral images during the degradation process. Employing a constant time interval is not effective, as setting it too short may lead to inefficiencies within the system while opting for a long interval can result in missing critical data. Moreover, the rate of polymer degradation can vary throughout the course of an experiment, adding to the complexity of the task. To address these challenges, we have developed a sequential sampling algorithm, as shown in Alg. \ref{alg: time_opt}, designed for optimising the time intervals between hyperspectral images. The process is as follows:

Firstly, each sample's total reflected intensity ($I$) is extracted from the ROI of hyperspectral images and recorded after each measurement. Next, a subset of these intensities, usually the last two measurements and a desired rate of change, denoted as $c$, are used to calculate the ideal change in intensity ($i_e$). Any change smaller than a predefined degradation limit ($\epsilon$) is considered insignificant. If the ideal change in intensity is smaller than the defined limit $\epsilon$ (e.g. $<1\%$ of the starting intensity), it can be inferred that the sample won't undergo any further notable degradation, and the experiment can be stopped. Alternatively, sampling can continue without adjusting if the difference between $i_e$ and the last measured change in intensity ($\Delta i$) is within an acceptable margin ($\delta$).

However, an optimisation algorithm is utilised if none of these conditions are met. This algorithm calculates the optimal time for the next measurement ($t_{n+1}$) by considering several factors. First, it computes a normalised distance ($d$) between the ideal and actual change in intensity, with the option to increase sensitivity by applying a gain factor ($g$). A simple way of calculating the normalised distance between $i_e$ and $\Delta i$ is given in Eq. \ref{eq: sss_norm}.

\begin{equation}
    \label{eq: sss_norm}
    d = g \cdot (\frac{\Delta i}{i_e} - 2 + \frac{i_e}{\Delta i})
\end{equation}

 To find the appropriate change in sampling time ($s$), we first need to define the maximum allowed factor of change per time interval $f$ (we usually don't want this to be more than a factor of two). We then introduce a smooth monotone mapping of distances $d$ to $s$, using the $f$ factor as given by Eq. \ref{eq: sss_map}. 

\begin{equation}
    \label{eq: sss_map}
    s = (1-f) e^{-d} + f
\end{equation}

Finally, $t_{n+1}$ is determined based on the elapsed between the two previous measurements $\Delta t$, $s$, and absolute lower ($b_l$) and upper ($b_u$) time boundaries. These absolute boundaries are established based on system limitations and the specific requirements of the experiment. If there are multiple samples imaged, the final $t_{n+1}$ that the robotic system uses is the smallest $t_{n+1}$ between all the given samples. The proposed Alg. \ref{alg: time_opt} offers notable advantages, primarily stemming from its independence regarding how intensities change over time and its low computational cost. These characteristics make it a flexible and efficient tool for optimising the timing of hyperspectral image capture during experiments.

%%%%%%%%%%%%%%%%%%%%%%%%%%%%%%%%%%%%%%%%%%%%%%%%%%%%%%%%%%%%%%%%%%%%%%%%%%%%%

\section{Experimental setup and results}
\label{result sections}
\subsection{Apparatus and integration}

A Universal Robotics UR5e collaborative robot was selected as the robotic arm and was equipped with an OnRobot RG6 gripper. Custom detachable 3D-printed fingers were implemented for specific applications in the hand.

The SPECIM FX10e push-broom hyperspectral camera serves as the primary characterisation device. It provides a spatial resolution of 1024 pixels and covers the wavelength range of 400 – 1000nm with a spectral resolution of 224. The main light source for hyperspectral imaging is a 500W linear halogen lamp.

ROS1 Noetic on Ubuntu 20.04 and standard UR5e and RG6 gripper drivers were utilised for system integration and communication. The robotic arm was controlled by directly using its controllers and the Transform Library \cite{6556373} for path planning.

One of the main challenges in capturing hyperspectral data was ensuring no information loss and maintaining 16-bit depth through multiple software frameworks. The GigE interface of the FX10e limited the camera driver's choices. Aravis was the preferred driver option and was used as an element within a Gstreamer pipeline, allowing the transport of data to be fine-tuned and previewed.

OpenCV \cite{opencv_library} was used to manage the Gstreamer pipeline and save the output to a cloud server. It’s important to note that the above process was interfaced with the experiment sequencer, allowing the automatic capture of hyperspectral images in concert with the robot's motion.

\subsection{Polymer degradation experimental procedure}
An example of a sample preparation procedure we undertook is as follows: Polymer (5 mg) was dissolved in 2 mL of 1,4-dioxane, 400 \textmu L was taken and added dropwise to a 15 mL centrifuge tube that contains 8 mL of a solution of 99\% ethanol and 1\% normal saline (0.9\% NaCl in water). The mixture was allowed to settle for at least 30 minutes, then it was centrifuged at 3500 RPM for 1 minute, and the supernatant was removed. The solid was then dispersed in 1.5 mL of PBS at 0.6M concentration and pH 12.3 (unless mentioned otherwise). This suspension was then dispensed into a 24 well plate ($\rho A$ = 2.63 $mg/cm^2$). Through several experiments, we find that the mass of polymer per imaged-surface-area ($\rho A$), 3 $> \rho A >$ 2 $ mg/cm^2$ provides the best results for our given polymers and experimental setup. The sample well plate was then put on an orbital shaker, shaking at 30 RPM. The sequencer program was launched, and the tasks defined in Fig. \ref{fig: img_proc} were executed. The defined process was repeated at specified time intervals for a defined number of cycles, depending on the experiment's requirements.

\begin{figure}[t]
    \centering
    \includegraphics[width=1\columnwidth]{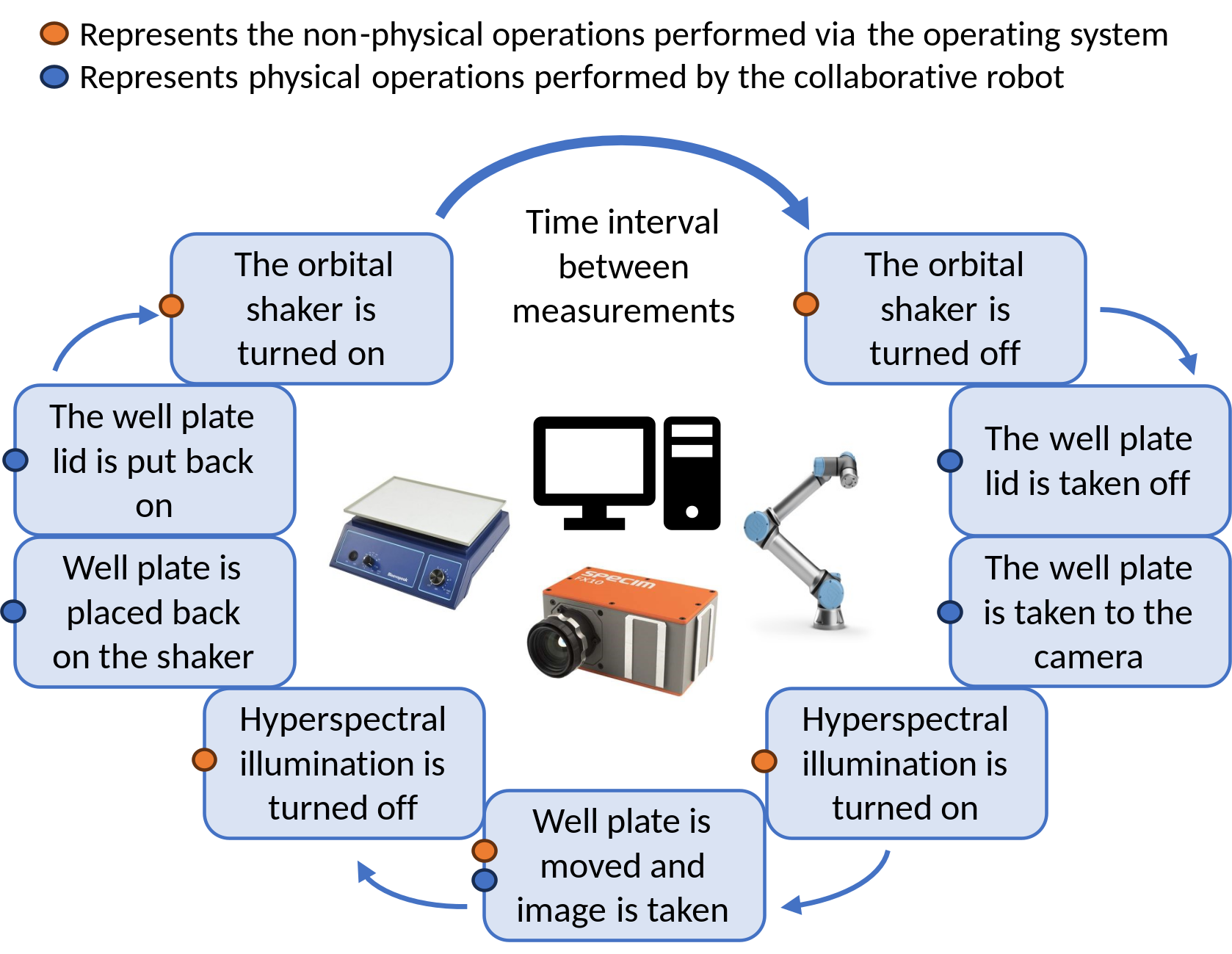}
   \caption{Schematic diagram of the imaging procedure. The lid manipulations are crucial steps to minimise evaporation when idle and mitigate any potential artefacts during the imaging process.}
    \label{fig: img_proc}
\end{figure}

Table \ref{tab:poly_info} contains information about different polymers used in our experiments.

\begin{table}[!t]
    \caption{Polymer information}
    \label{tab:poly_info}
    \centering
    
    \begin{tabular}{|c|c|c|c|c|c|}
        \hline
         Name  & Mn & Mw & Endcap & Dispersity & Source \\
         \hline 
         \hline
         PLA1 & 56687 & 82911 & Ester & 1.463 & Natureworks \\
         \hline
         PLA2 & 200913 & 296975 & Acid & 1.478 & Akina Polymers\\
         \hline 
         PLA3 & 11075 & 15089 & Ester & 1.362 & Akina Polymers\\
         \hline 
         PCL1 & 19006 & 34345 & Acid & 1.807 & Merck (Sigma)\\
         \hline
         PHB1 & 128766 & 284529 & None & 2.210 & Merck (Sigma)\\
         \hline
    \end{tabular}
    
\end{table}

\subsection{Uncertainty and performance}
Each measurement conducted during any experimental run exhibited a relative standard deviation of 4.05\%. This was calculated by putting identical samples within the wells and comparing their differences. The given standard deviation accounts for the systematic error inherent in our designed system. Furthermore, when considering both PLA1 and PLA2, the standard deviation of half-life measurements within the same experimental run was determined to be approximately 21 minutes. As a result, we can assert a 95\% confidence level in the data's self-consistency within a margin of error of 42 minutes for all measured half-life values pertaining to PLA1 and PLA2 within the same run. It's worth noting that this margin could potentially be wider for polymers with longer degradation periods or measurements made among different runs.
The circle detection algorithm exhibits a self-optimising capability, leading to 100\% accuracy in detecting regions of interest (ROIs) for well plates of size 24 or smaller, contingent upon providing reasonable initial values to the algorithm.
The final iteration of the robotic system achieved an independent success rate of 98.75\% per measurement. This entailed a total of 396 successful measurements collected and analysed, with only 5 instances resulting in system failure (likely to be the result of computer glitches). Notably, the system's longest consecutive successful run comprised 156 measurements, spanning a week-long period without human intervention.

\subsection{Performance of degradation models}
\label{perf_deg_model_ss}

\begin{figure}[t]
     \centering
     \begin{subfigure}[b]{0.89\columnwidth}
        \centering
        \includegraphics[width=1\columnwidth]{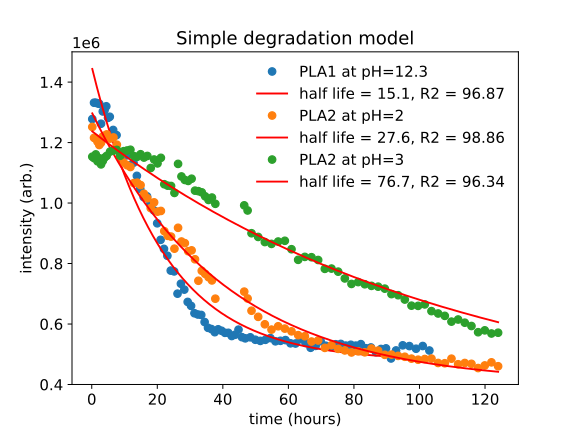}
        \caption{accuracy ($R^2$ values) of the exponential degradation model}
        \label{fig: simple_deg}
     \end{subfigure}
     \hfill
     \begin{subfigure}[b]{0.89\columnwidth}
        \centering
        \includegraphics[width=1\columnwidth]{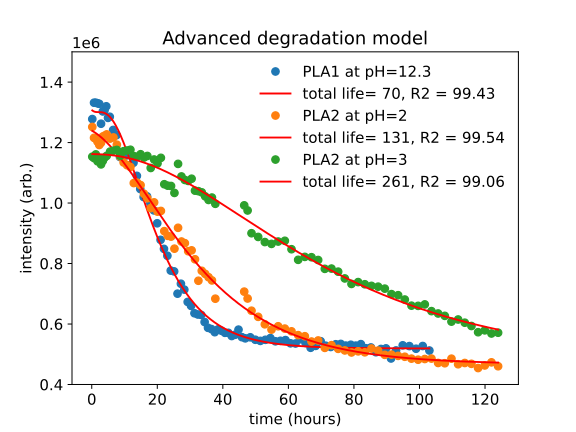}
        \caption{accuracy ($R^2$ values) of the advanced degradation model}
        \label{fig: adv_deg}
     \end{subfigure}
     \hfill
     \caption{Degradation model-fitting performance with fixed sampling rate. \vspace{-0.5cm}}
        \label{fig: deg_model_seq_samp1}
\end{figure}

Two distinct mathematical models (Eq. \ref{eq:final_decay} and Eq. \ref{eq:new_model_solved}) were formulated to describe the behavior of PLA in a PBS when utilising hyperspectral imaging. As illustrated in Figures \ref{fig: simple_deg} and \ref{fig: adv_deg}, the model incorporating a linear combination of both surface area and mass-dependent degradation rates demonstrates superior performance, achieving a $R^2$ value of over 99\% for the examined polymers. In comparison, the simpler exponential model attains $R^2$ values in the range of approximately 96-98\%. This eludes the accuracy of the developed model for describing the factors underlying polymer degradation.
However, it's worth noting that the advanced surface-mass-dependent degradation model doesn't have a half-life as an inherent characteristic. Consequently, we introduce a new metric called ``total life", signifying the time required for the hyperspectral signal to diminish to less than 1\% of its initial amplitude, directly translating to more than 99\% of the polymer degraded and dissolved into the PBS. Table \ref{tab:model_perf} compares the performance of the two different models using AIC and MDL; the maximum value data point was normalised to $1$ prior to applying the model selection metrics. For all three of the given samples, the surface-mass-dependent degradation model (denoted as \textit{adv.}) has a lower AIC and MDL, indicating a more descriptive model for the degradation process compared to a simple exponential approach (denoted as \textit{exp.}).
Despite the fact that the surface-mass-dependent model provides a more accurate description of polymer degradation, we will use the exponential model (Eq. \ref{eq:final_decay}) for subsequent analyses to report the well-established concept of half-life. This is justified as the $R^2$ values fitting the exponential model fall within an acceptable range.

\begin{table}[!t]
    \caption{Degradation model performance}
    \label{tab:model_perf}
    \centering
    \begin{tabular}{|c|c|c|c|c|}
        \hline
         Sample & AIC(exp.) & AIC(adv.) & MDL(exp.) & MDL(adv.) \\
         \hline 
         \hline
         PLA1 at pH=12.3 & 0.14 & \textbf{0.026} & 0.32 &  \textbf{0.070}  \\
         \hline
         PLA2 at pH=2 & 0.048 & \textbf{0.020} & 0.11 & \textbf{0.054} \\
         \hline 
         PLA2 at pH=3 & 0.072 & \textbf{0.026} & 0.17 & \textbf{0.072} \\
         \hline
    \end{tabular}
\end{table}

\begin{figure}[t]
     \centering
     \begin{subfigure}[b]{0.89\columnwidth}
        \centering
        \includegraphics[width=1\columnwidth]{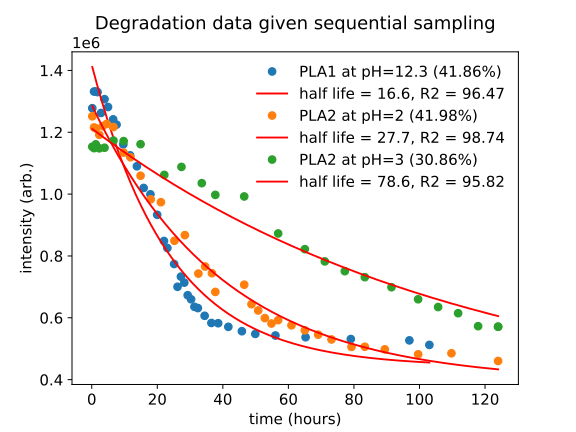}
        \caption{Sequential sampling algorithm and exponential model fitting}
        \label{fig: opt_exp_decay}
     \end{subfigure}
     \hfill
     \begin{subfigure}[b]{0.89\columnwidth}
        \centering
        \includegraphics[width=1\columnwidth]{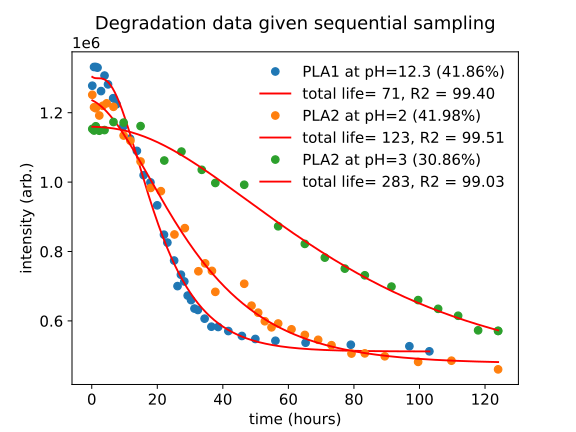}
        \caption{Sequential sampling algorithm and advanced model fitting}
        \label{fig: opt_adv_deg}
     \end{subfigure}
        \caption{Performance of model fitting via variable sequential sampling algorithm. The $R^2$ value given is between the fitted model using the post-sampling data and the original data set. The percentage of data used after sequential sampling is given within the brackets}
        \label{fig: deg_model_seq_samp2}
\end{figure}

\subsection{Performance of variable sequential sampling strategy}

The sequential sampling strategy is applied to existing data to assess its performance. The objective is to evaluate the trade-off between the reduction in the number of data points and the resulting measurement accuracy. In practice, the algorithm would predict the optimal time for the next measurement and then select the closest available data point to that predicted time. When comparing Figures \ref{fig: opt_adv_deg} and \ref{fig: opt_exp_decay} to \ref{fig: adv_deg} and \ref{fig: simple_deg}, respectively, a notable reduction in the number of data points used is observed, down to only 30-42\% of the original data set. Despite this reduction in the number of data points, the measured half-life and total life remain within 10\% of the values obtained using the full data set. This demonstrates the efficiency and effectiveness of the sequential sampling strategy in optimising data collection while maintaining measurement accuracy. The algorithm's performance is assessed offline using the collected data in section \ref{perf_deg_model_ss}. This approach enables a direct comparison between sampling methods, focusing exclusively on the distinction between fixed and time-variant intervals while minimising the influence of other variables.

\subsection{Additional polymer types}
\label{not_limited_to_pla}
Throughout our experimentation, PLA served as the primary polymer due to its relatively rapid degradation, enabling us to explore and demonstrate the capabilities of the characterisation system. However, as illustrated in Fig. \ref{fig: phb_pcl_run}, the system showcases its capabilities beyond PLA by successfully measuring the half-life of other polymers, such as PHB and PCL.

\begin{figure}
    \centering
    \includegraphics[width=0.85\columnwidth]{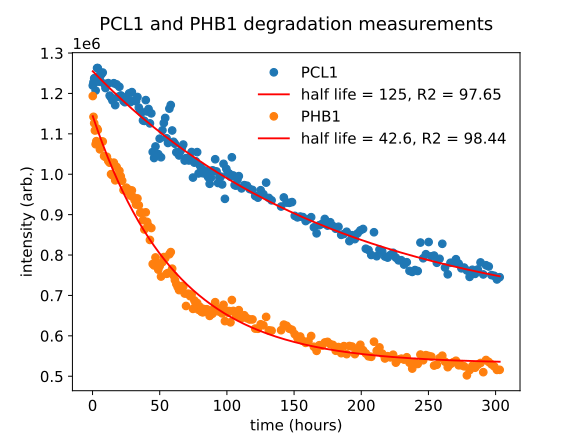}
    \caption{The degradation of PCL1 and PHB1 were measured over the course of two weeks. PBS of strength pH 14 was used to attain these results.\vspace{-0.4cm}}
    \label{fig: phb_pcl_run}
\end{figure}

\begin{figure}
    \centering
    \includegraphics[width=0.85\columnwidth]{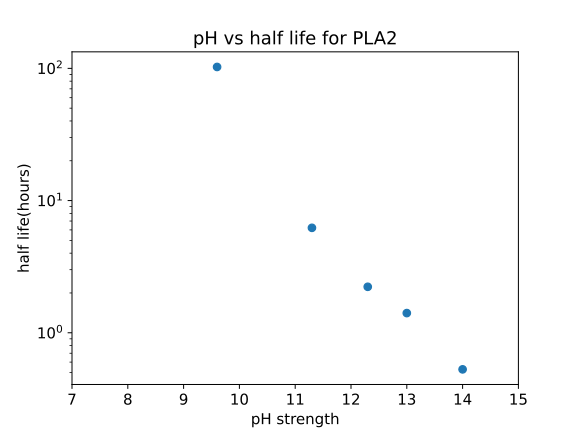}
    \caption{Illustrates the effect of pH strength on the half-life of PLA2}
    \label{fig: ph_vs_hl}
\end{figure}

\subsection{Effect of pH}
The degradation of PLA at different pH strengths was measured to give a quantitative analysis of the effect of pH strength on the polymer degradation rate. As demonstrated in Fig. \ref{fig: ph_vs_hl}, increasing the pH of PBS from 9.6 to 12.3 significantly accelerated polymer degradation, reducing the half-life of PLA2 by almost 50 times. However, it's important to note that this effect could vary among different polymers and buffer solutions. Hence, our system provides the methodology needed for high-throughput degradation measurement under various conditions. This methodology is not limited to the variability of pH strength and can be extended to polymer mass, temperature, buffer type or concentration.

\subsection{GPC measurements}
The samples undergoing degradation were first imaged and then periodically taken out of the wells at specific time intervals to conduct the destructive GPC analysis. This step was crucial to confirm that the changes observed by the hyperspectral camera were caused by the polymer degradation process. As depicted in Fig. \ref{fig: whole_gpc}, the sample at t=0, which had not undergone any degradation, displayed a broad peak with a central value of Mn $\approx$ 200,000 g/mol. At each subsequent time point, we observe the original peak diminishing in intensity, while two additional peaks emerge at Mn $\approx$ 2,500 and Mn $\approx$ 5,000 g/mol. Notably, no polymer is detected at Mn $<$ 1,000 g/mol. This absence in the lower Mn range could be attributed to the partial dissolution of oligomers below this molecular weight threshold, resulting in rapid degradation with only monomers remaining.

Fig. \ref{fig: hyperspec_gpc} presents the hyperspectral measurements corresponding to the GPC results, which are illustrated earlier in Fig. \ref{fig: whole_gpc}. Within a span of 25 hours, it appears that the polymer has undergone near-complete degradation, with the exponential model closely approaching its projected asymptote. Consequently, Fig. \ref{fig: gpc_5_vs_25} showcases the final GPC measurement acquired at the 25th hour compared to the 5th hour of the experiment. This measurement shows a minute peak at Mn $\approx$ 2,200 g/mol. This observation serves as additional confirmation, reinforcing the accuracy and reliability of the degradation measurements conducted by our proposed system.

\begin{figure}[t]
     \centering
     \begin{subfigure}[]{0.85\columnwidth}
        \centering
         \adjustbox{trim=0cm 0cm 0.5cm 0cm, clip}{\includegraphics[width=1\columnwidth]{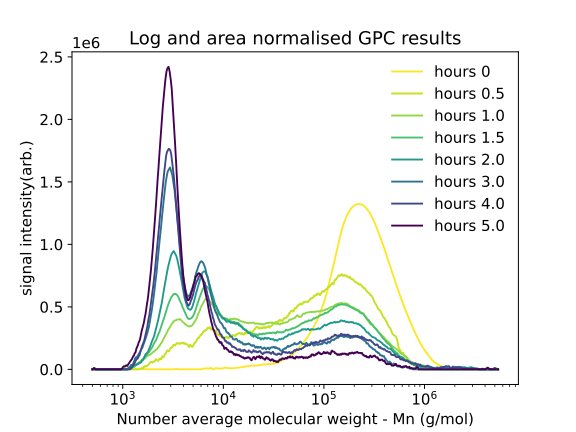}}
        \caption{}
        \label{fig: whole_gpc}
     \end{subfigure}
     %\hfill
     \begin{subfigure}[]{0.85\columnwidth}
        \centering
        \adjustbox{trim=0cm 0cm 0.5cm 0cm, clip}{\includegraphics[width=1\columnwidth]{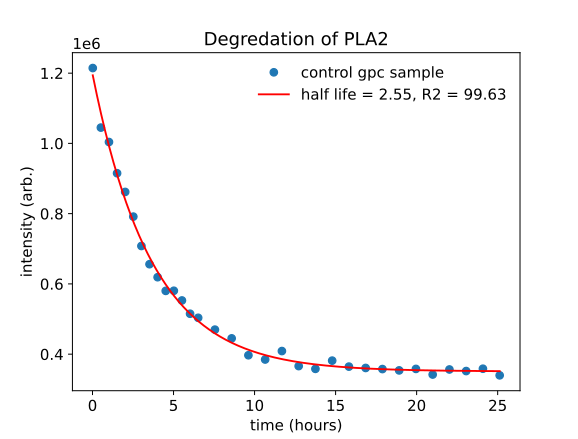}}
        \caption{}
        \label{fig: hyperspec_gpc}
     \end{subfigure}
     %\hfill
     \begin{subfigure}[]{0.85\columnwidth}
        \centering
         \adjustbox{trim=0cm 0cm 0.5cm 0cm, clip}{\includegraphics[width=1\columnwidth]{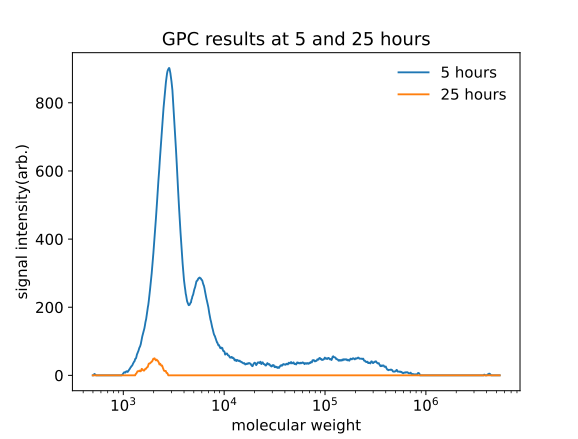}}
        \caption{}
        \label{fig: gpc_5_vs_25}
     \end{subfigure}
        \caption{(a) A plot of the GPC measurements over the course of 5 hours. (b) The hyperspectral results of the control sample in the GPC run. (c) The GPC measurements at 5 hours and 25 hours time points.}
        \label{fig: gpc_results}
\end{figure}

\subsection{Effect of molecular weight and end-groups on degradation}
Blends of three pairs of polymers (PLA1-PLA2, PLA1-PLA3 and PLA2-PLA3) with different molecular weights were made, resulting in a range of samples with varying number average molecular weights (Mn). The half-life of these blends was measured and is illustrated in Fig. \ref{fig: mn_vs_halflife}. The results demonstrate that PLA1 has the highest half-life despite having a smaller Mn than PLA2. Additionally, we know that PLA1 and PLA3 are both ester end-capped while PLA2 is acid end-capped. These results show that the degradation rate is affected by a multitude of inter-dependant factors, such as the combination of end groups and molecular weight. Additionally, the convergence of half-life measurements between two distinct sets of blends (PLA1-PLA2 and PLA1-PLA3) and measurements show the accuracy of the measured half-life.

\subsection{Complex blends}
To fully demonstrate the capabilities of the proposed system, we synthesised a blend comprising three distinct polymers (PLA1-PLA2-PLA3) and subsequently measured their respective half-lives, as shown in Fig. \ref{fig: 3D ternary plot}. Many variables affect the degradation rate of polymers; hence, a high throughput analysis of a wide range of polymers and their blends through an automated process is crucial.

\begin{figure}
    \centering
    \includegraphics[width=1\columnwidth]{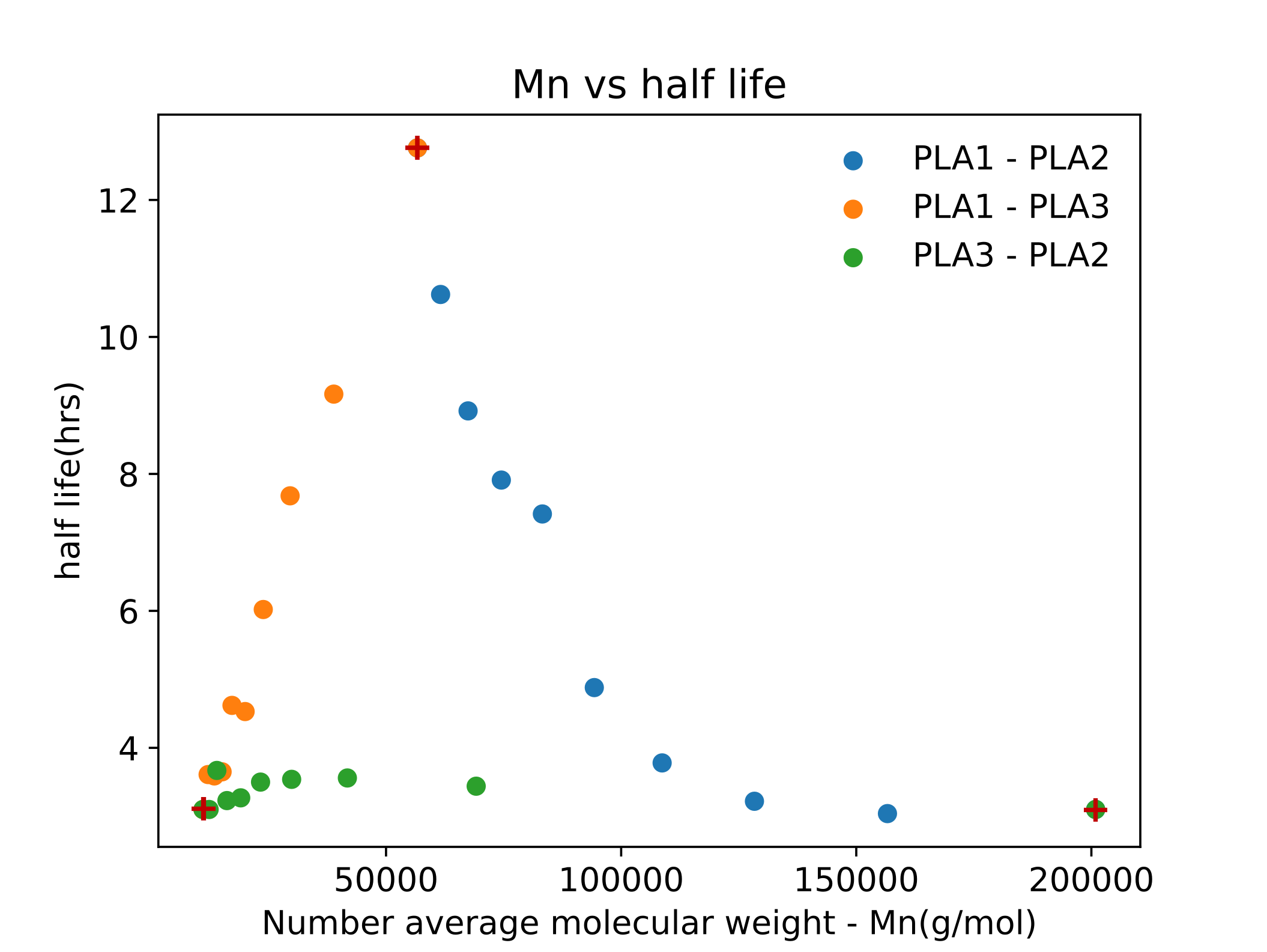}
    \caption{Blend of pairs of polymers demonstrating the effect of molecular weight and end groups. The ``+'' on a data point represents that it's a shared point between two blends.}
    \label{fig: mn_vs_halflife}
\end{figure}

\begin{figure}[t]
    \centering
    \includegraphics[width=1\columnwidth]{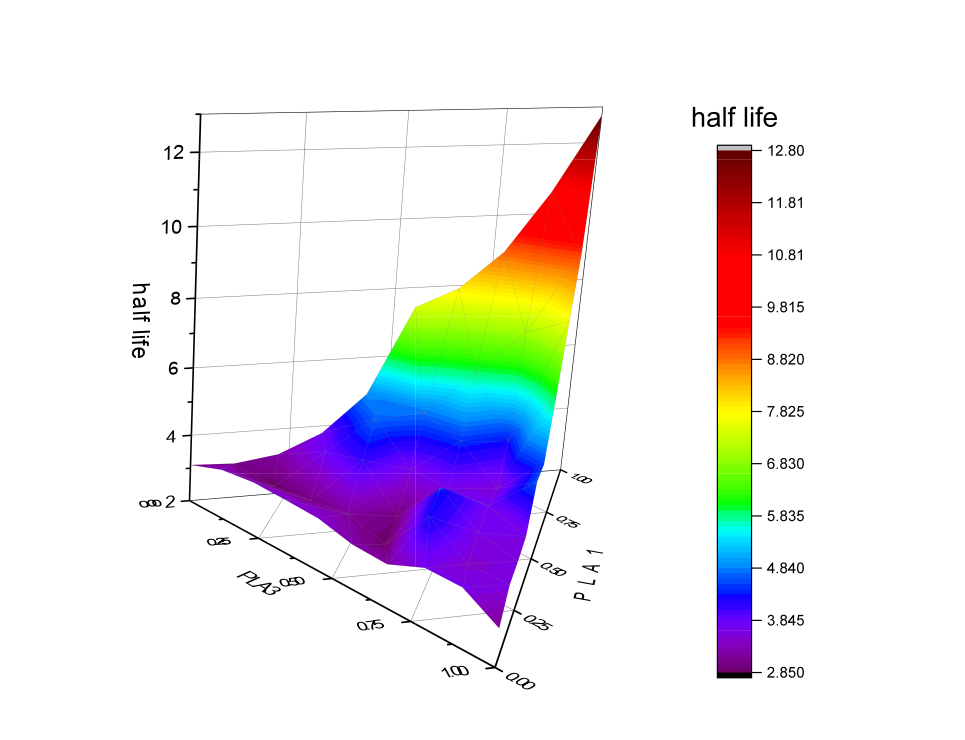}
    \caption{3D ternary plot of the half life of PLA1, PLA2, PLA3 and their blends}
    \label{fig: 3D ternary plot}
\end{figure}

\section{Discussion}
The characterisation station successfully measured the half life of staple biodegradable polymers such as PLA, PHB, and PCL within a laboratory environment. The GPC measurements confirm the observed change in hyperspectral images is due to polymer degradation followed by dissolution. Due to an increase in pH strength, the acceleration of degradation was measured quantitatively to be a factor of 2-5 per integer increase in pH within the range of 9.6 and 14 for PLA2. To fully quantify the extent of change in the degradation rate due to an increase in pH strength, a more extensive study analysing a variety of polymers will be required. 

Furthermore, by utilising blends of PLA1, PLA2 and PLA3, we identified a gradual but non-linear change in the half-life of the PLA samples when they were blended at different ratios. This is critical as without an extensive study of their blends, some results could be interpreted as outliers. The ternary plot in Fig. \ref{fig: 3D ternary plot} depicts the smooth transition of half-life between the different polymers to verify the genuine effects of molecular weight and end groups underpinning the degradation rate. This lays the groundwork for automated and high throughput analysis of polymer blends and their properties within the realm of material science.

The sequential sampling strategy developed to optimise the time interval between hyperspectral measurements can be implemented in real-time as a next step to improve the efficiency and multi-experimental throughput of the system. Additionally, computationally efficient machine learning algorithms could be explored as an alternative method to the presented method. All the results provided in this paper were acquired individually, leading to extended idle periods of the characterisation station. However, the system is capable of running multiple independent or dependent experiments in parallel, which should be considered in the future use of the developed system for other applications.

\section{CONCLUSION}
An innovative robot-assisted automated hyperspectral imaging-based methodology is proposed to facilitate the temporal characterisation of materials. Through the utilisation of this developed system, hyperspectral analysis is introduced as a novel technique for evaluating the degradation rate of biodegradable polymers. 
The proposed automated system has accurately measured polymer degradation rates at high throughput. This capability enables the characterisation of degradation rates across a wide range of different conditions. Notably, our observations indicate that multiple factors influence degradation rates, with end groups emerging as the most influential polymer property. Our work lays the foundation for further exploration through in-depth investigations into the individual factors, ultimately expanding our collective understanding of polymer degradation dynamics. The proposed system is shown to be capable of generating systematic data for the validation of models. Future work could utilise this capability to test the suitability of potential models for material characterisation. There exists ample potential for extending the application of our demonstrated system to a broader range of temporal experiments encompassing various materials and conditions.

%%%%%%%%%%%%%%%%%%%%%%%%%%%%%%%%%%
\addtolength{\textheight}{-5 cm}

\bibliographystyle{IEEEtran.bst}
\bibliography{robopolyscan.bib}
%%%%%%%%%%%%%%%%%%%%%%%%%%%%%%%%%

\end{document}